\documentclass{aa}
\usepackage{graphicx}
\usepackage{txfonts}
\usepackage{epsfig,latexsym,longtable,lscape}

\usepackage{natbib}
\bibpunct{(}{)}{;}{a}{}{,}

\def\xmm{XMM-Newton}
\newcommand{\nh}{N$_{\rm H}$}
\newcommand{\lx}{\hbox{L$_{\rm x}$}}

\newcommand{\fx}{\hbox{F$_{\rm x}$}}
\newcommand{\expo}[1]{$\times 10^{#1}$}
\newcommand{\oexpo}[1]{$10^{#1}$}

\newcommand{\hcm}[1]{$\times 10^{#1}$ cm$^{-2}$}

\def\HII{\hbox{H\,{\sc ii}}}
\newcommand{\SII}{[S\,{\sc ii}]}
\newcommand{\OIII}{[O\,{\sc iii}]}
\newcommand{\Halpha}{H$_{\alpha}$}
\DeclareRobustCommand{\ion}[2]{%
\relax\ifmmode
\ifx\testbx\f@series
{\mathbf{#1\,\mathsc{#2}}}\else
{\mathrm{#1\,\mathsc{#2}}}\fi
\else\textup{#1\,{\mdseries\textsc{#2}}}%
\fi}\newcommand{\D}{$^\circ$}
\def\p0{\phantom{0}}

\def\it{\sl}

\begin{document}

   \title{New XMM-Newton observations of SNRs in the SMC}
   \author{M.D.~Filipovi\'c\inst{1,2}
            \and F.~Haberl\inst{1}
            \and P.F.~Winkler\inst{3}
            \and W.~Pietsch\inst{1}
            \and J.L.~Payne\inst{4}
            \and E.J.~Crawford\inst{2}
            \and A.Y.~De~Horta\inst{2}
            \and F.H.~Stootman\inst{2}
            \and B.E.~Reaser\inst{5}
           }

    \institute{Max-Planck-Institut f\"ur extraterrestrische
    Physik, Giessenbachstrasse, D--85741 Garching, Germany\\
    e-mail: fwh@mpe.mpg.de, wnp@mpe.mpg.de
    \and
    University of Western Sydney, Locked Bag 1797, Penrith South DC,
    NSW 1797, Australia\\
        e-mail: m.filipovic@uws.edu.au, e.crawford@uws.edu.au,
        a.dehorta@uws.edu.au, f.stootman@uws.edu.au
      \and
    Department of Physics, Middlebury College, Middlebury, VT 05753,
    USA\\
        e-mail: winkler@middlebury.edu
    \and
    Centre for Astronomy, James Cook University, Townsville QLD,
4811, Australia\\
        e-mail: Jeffrey.Payne@jcu.edu.au
     \and
     Department of Physics and Astronomy, Swarthmore College, 500 College Avenue, Swarthmore, PA 19081, USA
       }

  \abstract
    {A complete overview of the supernova remnant (SNR) population is required to investigate their 		evolution and interaction with the surrounding interstellar medium in the Small Magellanic
    Cloud (SMC).}
    {Recent XMM-Newton observations of the SMC cover three known SNRs (DEM\,S5, SNR\,B0050$-$72.8, and 	SNR\,B0058$-$71.8), which are poorly studied and are X-ray faint.
    We used new multi-frequency radio-continuum surveys and new optical
    observations at \Halpha, \SII, and \OIII\ wavelengths, in combination with the X-ray data, to 		investigate their properties and to search for new SNRs in the SMC.}
    {We used X-ray source selection criteria and  found one SMC object with typical SNR
    characteristics (HFPK\,334), that was initially detected by ROSAT.
    We analysed the X-ray spectra and present multi-wavelength morphological studies
    of the three SNRs and the new candidate.}
    {Using a non-equilibrium ionisation collisional plasma model, we find
    temperatures kT around 0.18~keV for the three known remnants and 0.69~keV for
    the candidate. The low temperature, low surface brightness, and large extent
    of the three remnants indicates relatively large ages. The emission from the
    new candidate (HFPK\,334) is more centrally peaked and the higher temperature suggests
    a younger remnant. Our new radio images indicate
    that a pulsar wind nebulae (PWN) is possibly associated with this object.}
    {The SNRs known in the SMC show a variety of morphological structures that are
    relatively uncorrelated in the different wavelength bands, probably caused by the 
    different conditions in the surrounding medium with which the remnant interacts.}

\keywords{Galaxies: Magellanic Clouds -- ISM: supernova remnants
 -- individual: DEM\,S5, SNR\,B0050$-$72.8, SNR\,B0058$-$71.8,
HFPK\,334}
%
\maketitle
%

\section{Introduction}

The study of X-ray supernova remnants (SNRs) in nearby galaxies is
of major interest for understanding both the X-ray output of more
distant galaxies and the processes that occur on local
interstellar scales within our own Galaxy. Unfortunately, the
distances to many Galactic remnants are uncertain by a factor of 2,
leading to a factor of 4 uncertainty in luminosity and of 5.5 in the
calculated energy release of the initiating supernova (SN). At a
distance of $\sim$60~kpc \citep{2005MNRAS.357..304H}, the Small
Magellanic Cloud (SMC) is one of the prime targets for the
astrophysical research of extragalactic objects, including SNRs. This is
because these remnants are located at a known distance, yet close
enough to allow a detailed analysis of them.

SNRs reflect a major process in the elemental enrichment of the
interstellar medium (ISM). Multiple supernova explosions close
together generate super-bubbles typically hundreds of parsecs in
extent. Both SNRs and super-bubbles are prime drivers controlling the morphology
and the evolution of the ISM. Their properties are therefore crucial
for a full understanding of the galactic matter cycle.

Previous X-ray surveys of the SMC have been undertaken using the {\sl
Einstein} \citep{1992ApJS...78..391W}, {\sl ROSAT}
\citep{2000A&AS..142...41H}, and {\sl ASCA}
\citep{2003PASJ...55..161Y} satellites. These surveys revealed
discrete X-ray sources and a large-scale diffuse emission. In
particular, the high sensitivity and the large field of view of the
{\sl ROSAT} PSPC instrument has provided the most comprehensive
catalogue of discrete X-ray sources, 517 in an area of $\sim$18
square degrees \citep{2000A&AS..142...41H}. It also has revealed the
existence of a hot thin plasma within the SMC ISM having
temperatures between $10^{6}$ and $10^{7}$~K
\citep{2002A&A...392..103S}.

The \xmm\ \citep{2001A&A...365L...1J} archive contains a number of 
observations in the direction of the SMC, and they mainly cover the bar 
and part of the eastern wing of the galaxy \citep[][Haberl et al., in preparation]{2007arXiv0712.2720H}. Typical limiting point-source luminosities of a 
few $10^{33}$~erg~s$^{-1}$ are reached with \xmm\ and extended objects like 
SNRs can be easily resolved\footnote{At a distance of 60~kpc, $\sim$1\arcsec\ 
subtends 0.29~pc.}.

Today a total number of eighteen classified SNRs are known in the
SMC, which are listed in Table~\ref{smcsnrs} \citep[][and references
therein]{2005MNRAS.364..217F,2007MNRAS.376.1793P,2004A&A...421.1031V,2001AJ....122..849D,1984ApJS...55..189M,1982MNRAS.200.1007M,1994AJ....107.1363H,2004ApJ...608..208N}.
Thirteen of the remnants were covered by early \xmm\ observations
and a comprehensive study was presented by
\citet{2004A&A...421.1031V}. Archival data and our own proprietary
observations (from the AO5 and AO6 observing periods) cover three
more faint remnants. The inspection of the EPIC images also revealed a
new SNR candidate.

\begin{table*}
 \begin{center}
\tiny
 \caption[]{SNRs and SNR candidates in the SMC. The radio spectral index ($\alpha$)
 is taken from \citet{2005MNRAS.364..217F}
 and [S\,{\sc ii}]/H$\alpha$ ratio from \citet{2007MNRAS.376.1793P}.}
  \begin{tabular}{rlccccccl}
\hline\noalign{\smallskip}
\hline\noalign{\smallskip}
No& SNR Name  & R.A.       & Dec        & D        & S$_{\rm{1\,GHz}}$& $\alpha$& [S\,{\sc ii}]/H$\alpha$ & Other Names \\
  &           & (J2000)    & (J2000)    & [pc]             & [Jy]     &         &                         &       \\
\hline\noalign{\smallskip}
1 & DEM S5    & 00 41 00.1 &--73 36 30.0& 68.4$\times$62.6 & 0.057    &--0.9    & 0.8             & HFPK 530\\
2 & DEM S32   & 00 46 37.6 &--73 07 56.6& 40.7$\times$46.5 & 0.036    &--0.6    & 0.4             & Part of N S19, WW 15\\
3 & J0047-7306& 00 47 28.6 &--73 06 15.5& 32.0$\times$8.7  & 0.035    &--0.1    & 1.2             & part of N S19\\
4 & HFPK 419  & 00 47 36.5 &--73 09 14.0& 44.2$\times$33.7 & 0.215    &--0.6    & 0.4             & part of N S19, HFPK 419\\
5 & IKT 2     & 00 47 16.6 &--73 08 43.5& 29.1$\times$32   & 0.500    &--0.6    & 0.4             & N S19, DEM S32, WW 16, HFPK 413\\
\noalign{\smallskip}
6 & IKT 4     & 00 48 20.6 &--73 19 39.6& 45.1$\times$34.9 & 0.176    &--1.0    & 0.4             & N S24, WW 21, DEM S42, HFPK 454\\
7 & IKT 5     & 00 49 07.7 &--73 14 35.0& 48.3             & 0.034    &--0.6    & 0.7             & DEM S49, WW 22, HFPK 437\\
8 & IKT 6     & 00 51 06.7 &--73 21 21.4& 42.2             & 0.137    &--0.7    & 0.5             & 1E0049.4-7339, WW 24, HFPK 461\\
9 & B0050-72.8& 00 52 36.9 &--72 37 18.5& 42               & 0.273    &--1.0    & 0.4             & N S50, DEM S68, WW 30, HFPK 285\\
10& N S66D    & 00 58 00.0 &--72 11 01.4& 58.2             & 0.057    &\p00.0   & 0.4             &\\
\noalign{\smallskip}
11& IKT 16    & 00 58 17.8 &--72 18 07.4& 58               & 0.091    &--0.7    & 0.5             & WW 42\\
12& IKT 18    & 00 59 27.7 &--72 10 09.8& 40.7             & 0.501    &--0.8    & 1.7             & N S66, NGC 346, WW 44, HFPK 148\\
13& B0058-71.8& 01 00 23.9 &--71 33 41.1& 61               & 0.241    &--0.8    & \OIII           & DEM S108, HFPK 45\\
14& IKT 21    & 01 03 17.0 &--72 09 45.0& 20.9             & 0.123    &--0.5    & 0.5             & N S76C, WW 50\\
15& 1E0102-723& 01 04 01.2 &--72 01 52.3& 5.8$\times$10.5  & 0.363    &--0.6    & \OIII           & N S76, IKT 22, WW 51, HFPK 107\\
\noalign{\smallskip}
16& IKT 23    & 01 05 04.2 &--72 23 10.5& 55.9             & 0.112    &--0.7    & \OIII           & DEM S125, WW 52, HFPK 217\\
17& DEM S128  & 01 05 24.7 &--72 09 20.4& 37.8             & 0.060    &--0.5    & 0.6             & IKT 24, WW 53, HFPK 145\\
18& IKT 25    & 01 06 27.5 &--72 05 34.5& 32$\times$23     & 0.014    &--0.7    & 0.4             & DEM S131, WW 54, HFPK 125\\
\hline\noalign{\smallskip}
19 & HFPK 334  & 01 03 29.5 &--72 47 23.2& 17.5             & 0.025    &--0.5    & ---             & HFPK334\\
20 & B0113-729 & 01 13 33.8 &--73 17 04.4& 21.8             & 0.108    &--0.6    & 0.4?            & N S83C, DEM S147, HFPK 448\\
\noalign{\smallskip}\hline\noalign{\smallskip}
  \end{tabular}
 \end{center}
 \label{smcsnrs}
\end{table*}

Recently, radio-continuum observations at 20, 13, 6, and 3~cm (1420, 2370, 4800, and
8640~MHz) with the Australia Telescope Compact Array (ATCA) were
performed to study radio SNRs in the direction of the SMC
\citep[][and references therein]{2002MNRAS.335.1085F,
2005MNRAS.364..217F}. Follow-up optical spectroscopy allowed us to
confirm one candidate SNR and study 11 other known remnants. Line
intensity ratios provided rough estimates of the average SMC `metal'
abundance, supporting the idea that these ratios are related more to the
ISM than to SNR ejecta \citep{2007MNRAS.376.1793P}.

The present paper is organised as follows: Section
\ref{observations} outlines our selection and analysis of the \xmm\
data. We also discuss complementary radio and optical data in
greater detail. In Section \ref{results}, a multi-frequency analysis
of each remnant or candidate is presented and a discussion of SNRs
not detected in any X-ray observations. Our conclusions are given in
Section \ref{conclusions}.

\section{Observational Data}
 \label{observations}

\subsection{\xmm\ observations of SMC fields}

In order to investigate the X-ray source population of the SMC,
\citet{2007arXiv0712.2720H} started a systematic analysis of the
available \xmm\ data. This included
their own proprietary data together with data from the public
archive. Only observations performed with the EPIC-PN instrument
\citep{2001A&A...365L..18S} in {\sl Large Window} (LW), {\sl Full
Frame} (FF) or {\sl extended Full Frame} (eFF) imaging CCD readout
mode were considered. The data were analysed using the analysis
package XMMSAS version 7.0.0.

The \xmm\ observations cover three SNRs which have not yet been
studied in detail in X-rays due to their faintness: DEM\,S5
(HFPK\,530), SNR\,B0050$-$72.8 (DEM\,S68; HFPK\,285) and
SNR\,B0058$-$71.8 (DEM\,S108; HFPK\,45). All three were detected in
{\it ROSAT} PSPC data \citep[][with catalogue entry given by their
HFPK number]{2000A&AS..142...41H} and were not yet observed by \xmm\
at the time of the work of \citet{2004A&A...421.1031V}. Another {\it
ROSAT} PSPC source HFPK\,334, (with evidence for spatial extent in
the {\it ROSAT} data and a possible radio counterpart) shows X-ray
colours typical of SNRs and we suggest it as new candidate SNR.

In the following, we investigate the properties of the three X-ray
faint SNRs and the new candidate SNR HFPK\,334. For morphology
studies the \xmm\ EPIC images in different energy bands were
compared with radio-continuum and optical images (see below). For
X-ray spectral analysis EPIC spectra were extracted for PN (single +
double pixel events, corresponding to a PATTERN 0$-$4 selection) and
MOS \citep[][PATTERN 0$-$12]{2001A&A...365L..27T}, excluding bad CCD
pixels and columns (FLAG 0). These were fitted simultaneously,
allowing only a renormalisation factor to account for
cross-calibration uncertainties between the instruments.

We used {\sc XSPEC}\footnote{{http://heasarc.gsfc.nasa.gov/docs/xanadu/xspec/}} 
version 11.3.2p for spectral modelling. To
account for photo-electric absorption by interstellar gas, two
hydrogen column densities were used.
The first represents the foreground absorption in the Milky Way,
fixed at 6\hcm{20} assuming elemental abundances of
\citet{2000ApJ...542..914W}. The second considers the absorption in
the SMC \citep[with metal abundances set to 0.2 solar as typical of
the SMC; ][]{1992ApJ...384..508R}. The statistical quality of the
spectra was sufficient to fit one-component thermal plasma emission
models. We used a single-temperature non-equilibrium ionisation
collisional plasma (NEI) model \citep[in XSPEC, see][and references
therein]{2001ApJ...548..820B} with metal abundances fixed to 0.2
solar, yielding acceptable $\chi^2$ values.

\begin{figure*}
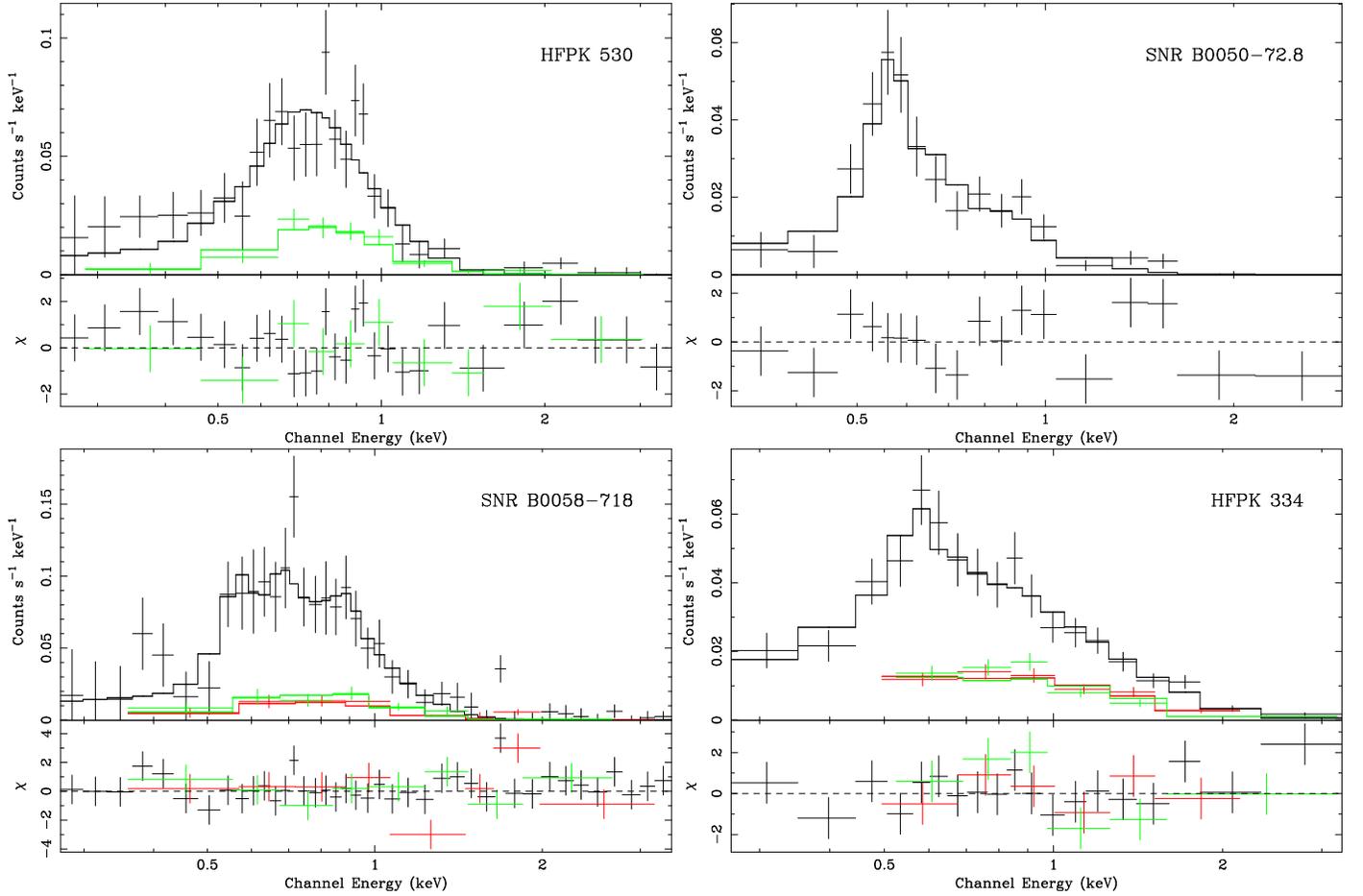

  \begin{center}
  \hbox{
  \resizebox{0.5\hsize}{!}{\includegraphics[angle=-90,clip=]{9642fig1a.ps}}
  \resizebox{0.5\hsize}{!}{\includegraphics[angle=-90,clip=]{9642fig1b.ps}}
  }
  \hbox{
  \resizebox{0.5\hsize}{!}{\includegraphics[angle=-90,clip=]{9642fig1c.ps}}
  \resizebox{0.5\hsize}{!}{\includegraphics[angle=-90,clip=]{9642fig1d.ps}}
  }
  \end{center}
  \caption{EPIC spectra of supernova remnants in the SMC. From top left to bottom right:
  a) SNR\,B0039$-$7353 = HFPK\,530, b) SNR\,B0050$-$72.8, c) SNR\,B0058$-$71.8 and
  d) the candidate SNR HFPK\,334. The best fits using a single-temperature NEI
  model are plotted as histograms (black: PN, red: MOS1, green: MOS2).}
  \label{fig-snr-spectra}
\end{figure*}

The EPIC spectra of all four of these X-ray sources are plotted in
Fig.~\ref{fig-snr-spectra} using their best-fit NEI models. The derived
best fit model parameters (SMC absorption, temperature kT and ionisation
time scale $\tau$) are summarised in Table~2 
together with inferred fluxes and luminosities. Flux and luminosity are given 
for 0.2--2 keV, determined from the EPIC-PN spectrum (HFPK\,530: MOS2). 
Intrinsic source luminosity with total \nh\ set to 0, assuming a distance of 
60~kpc to the SMC. The relatively large errors in the SMC absorption leads to 
large uncertainties in the (absorption corrected) luminosities. Therefore, we give
a luminosity range derived from fits with SMC \nh\ fixed at the lower
and upper confidence values. These very soft temperature estimates of 
\nh\ ranging from 7 to 12\expo{21} are consistent with very soft 
emission from older SNRs, obscured by heavy foreground extinction.
Similarly we derive a confidence range for the emission measure EM. 
The spectra of the three SNRs DEM\,S5, SNR\,B0050$-$72.8 and SNR\,B0058$-$71.8
are characterised by a high absorption and a low temperature.
Considering their large extent and low surface brightness, they are
most likely older remnants. The spectral analysis of the SNR
candidate HFPK\,334 yields a higher temperature, similar to what is 
found for most of the known SMC SNRs.


\begin{table*}
\begin{center}
\caption{Spectral fits to EPIC spectra of SNRs with an absorbed
NEI model. 
}
\begin{tabular}{lccccccccc}
\hline\noalign{\smallskip}
\hline\noalign{\smallskip}
\multicolumn{1}{l}{Source} &
\multicolumn{1}{l}{PN exp.} &
\multicolumn{1}{c}{SMC \nh} &
\multicolumn{1}{c}{kT} &
\multicolumn{1}{c}{$\tau$} &
\multicolumn{1}{c}{\fx} &
\multicolumn{1}{c}{\lx} &
\multicolumn{1}{c}{EM} &
\multicolumn{1}{c}{$\chi^2_r$} &
\multicolumn{1}{c}{dof} \\

\multicolumn{1}{c}{} & \multicolumn{1}{c}{[\oexpo{3} s]} &
\multicolumn{1}{c}{[\oexpo{21}cm$^{-2}$]} &
\multicolumn{1}{c}{[keV]} & \multicolumn{1}{c}{[\oexpo{8} s
cm$^{-3}$]} & \multicolumn{1}{c}{[erg cm$^{-2}$ s$^{-1}$]} &
\multicolumn{1}{c}{[erg s$^{-1}$]} & \multicolumn{1}{c}{[cm$^{-3}$]}
& \multicolumn{1}{c}{} &
\multicolumn{1}{c}{} \\

\noalign{\smallskip}\hline\noalign{\smallskip}
HFPK\,530         & 17.7 & 7.7$_{-1.9}^{+7.7}$ &  0.18$\pm$0.06  &  2$_{-1}^{+19}$  & 8.4\expo{-14} & 1.8$_{-0.9}^{+40} $ \expo{36} & 1.1$_{-0.5}^{+9.0}$ \expo{60} & 1.12 & 34 \\
SNR\,B0050$-$72.8 & 27.1 & 7.4$_{-2.4}^{+3.0}$ &  0.19$\pm$0.08  &  17$_{-6}^{+18}$ & 6.7\expo{-14} & 2.7$_{-1.8}^{+4.4}$ \expo{36} & 3.7$_{-2.7}^{+7.3}$ \expo{59} & 1.52 & 14 \\
SNR\,B0058$-$71.8 & 11.9 & 12.0$\pm$5.8        &  0.17$\pm$0.07  &  17$_{-13}^{+9}$ & 8.4\expo{-14} & 1.1$_{-1.0}^{+4.9}$ \expo{37} & 1.9$_{-1.6}^{+9.2}$ \expo{60} & 1.17 & 50 \\
HFPK\,334         & 17.7 & 2.7$\pm$0.9         &  0.69$\pm$0.12  &  4.8$\pm$1.2     & 1.3\expo{-13} & 3.6$_{-0.9}^{+1.3}$ \expo{35} & 3.1$_{-0.5}^{+0.5}$ \expo{58} & 1.19 & 25 \\
\noalign{\smallskip}\hline\noalign{\smallskip}
\end{tabular}
\end{center}
\label{tab-fits}
\end{table*}

\subsection{The ATCA radio-continuum surveys of the SMC}

We used all available radio-continuum images of the SMC, including
mosaics and specific ATCA target pointings. These are composed of
observations at 5 radio frequencies having moderate resolution at
36~cm (843~MHz) (MOST; \citet{1998PASA...15..280T}) and
20, 13, 6 and 3~cm (ATCA; \citet{2002MNRAS.335.1085F,
2001AJ....122..849D}). In addition, we use special lower-resolution
ATCA observations towards DEM\,S5 and SNR\,B0058$-$71.8 (DEM\,S108),
as described in \citet{2005MNRAS.364..217F}. These moderate
resolution observations ($>$12\arcsec) allow a reasonable insight
into the nature of these objects.

The ATCA observations of HPFK\,334 were made on 15$^{\mathrm{th}}$
November 2007 using array configuration EW367\footnote{http://www.narrabri.atnf.csiro.au/observing/configs.html}, and on 14$^{\mathrm{th}}$ January 2008 using array 
configuration 6A$^3$, at wavelengths of 6 and 3~cm. The observations
were done in so called `snap-shot' mode totalling $\sim$4 hours of
integration over a 12-hour period with each array. Source 1934-638 was used for
primary calibration and 0252-712 for secondary calibration. The
\textsc{miriad} \citep{2006miriad} and \textsc{karma}
\citep{2006karma} software packages were used for reduction and
analysis. At 6 and 3~cm, the image resolution is, respectively,
\mbox{21\arcsec $\times$ 20\arcsec} and \mbox{12\arcsec $\times$ 11\arcsec,} and the
r.m.s.\ noise is estimated to be 0.13~mJy/beam and 0.14~mJy/beam.

\subsection{The MCELS optical surveys of the SMC}
The Magellanic Cloud Emission Line Survey (MCELS) was carried out
from the 0.6 m University of Michigan/CTIO Curtis Schmidt telescope,
equipped with a SITE $2048 \times 2048$\ CCD, which gave a field of
1.35\degr\ at a scale of 2.4\arcsec\,pixel$^{-1}$. Both the LMC
and SMC were mapped in narrow bands corresponding to \Halpha,
\OIII\ ($\lambda$=5007\,\AA), and \SII\ ($\lambda$=6716,\,6731\,\AA),
plus matched red and green continuum bands that are used primarily
to subtract most of the stars from the images to reveal the full
extent of the faint diffuse emission. A $4.5\degr \times 3.5\degr$\
region, most of the SMC, was covered in 69 overlapping fields,
offset so that every point was included in at least four different
fields. All the data have been flux-calibrated and assembled into
mosaic images, small sections of which are shown in Figs.~2--4.
(Further details regarding the MCELS are given by \citet{2006NOAONL.85..6S},
Winkler et al. (in preparation) and at http://www.ctio.noao.edu/mcels/.)

\section{Results and discussion of individual objects}
 \label{results}

\subsection{The SMC SNR DEM\,S5 (HFPK\,530; J004100-733648)}

The emission nebulae DEM\,S5 \citep{1976MmRAS..81...89D} was
originally classified as an SNR candidate in our {\sl ROSAT} survey
\citep{2000A&AS..142...41H}. Optical spectroscopy and low-resolution
radio-continuum analysis of this object can be found in
\citet{2007MNRAS.376.1793P} and \citet{2005MNRAS.364..217F}. We
estimate DEM\,S5's radio spectral index\footnote{Defined as
$\alpha$, where $S_{\nu}\propto\nu^{\alpha}$} to be --0.9 and its optical
[S\,{\sc ii}]/H$_\alpha$ ratio about 0.8.

\begin{figure*} 
 \begin{center}
 \includegraphics[angle=-90,width=88mm]{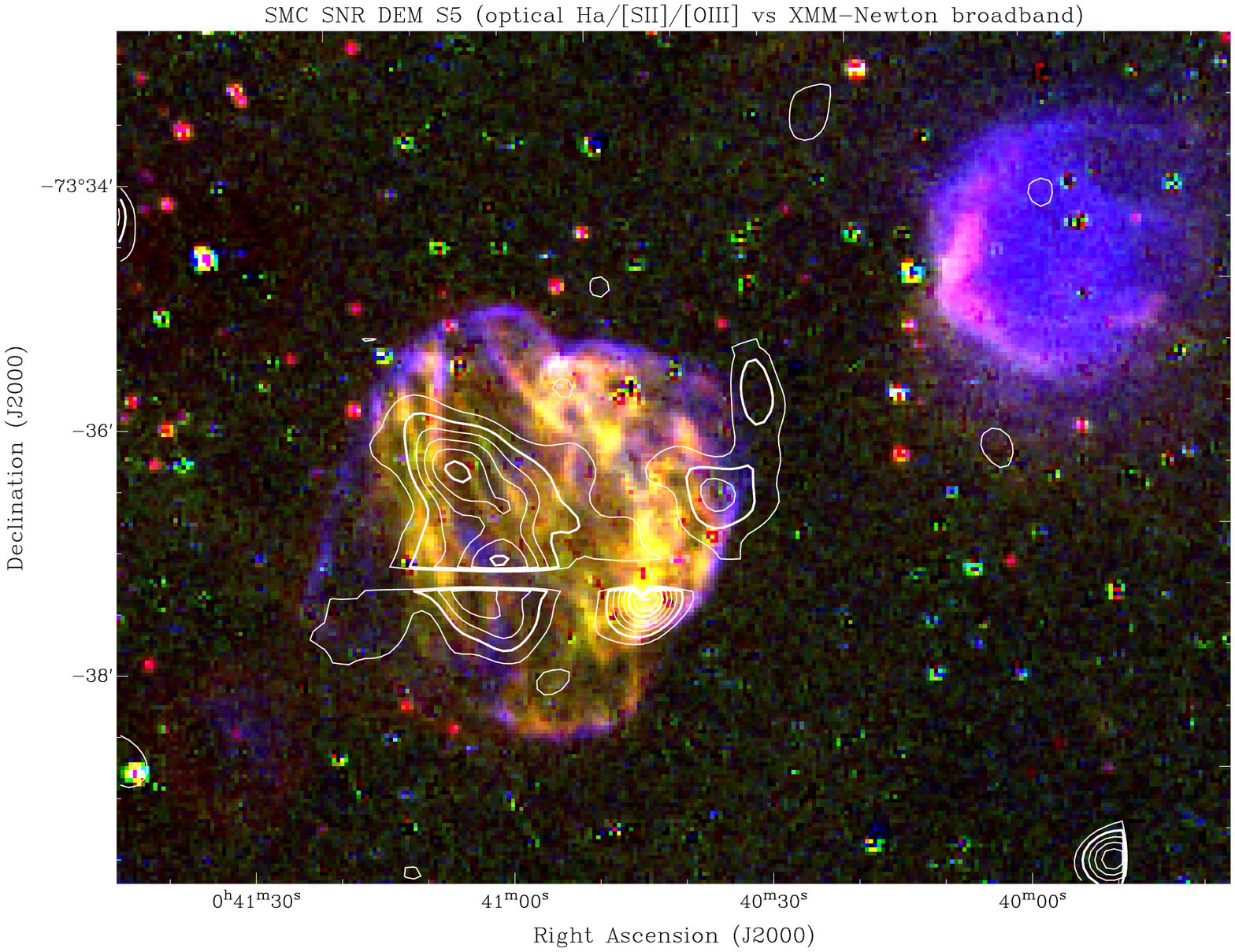}
 \includegraphics[angle=-90,width=88mm]{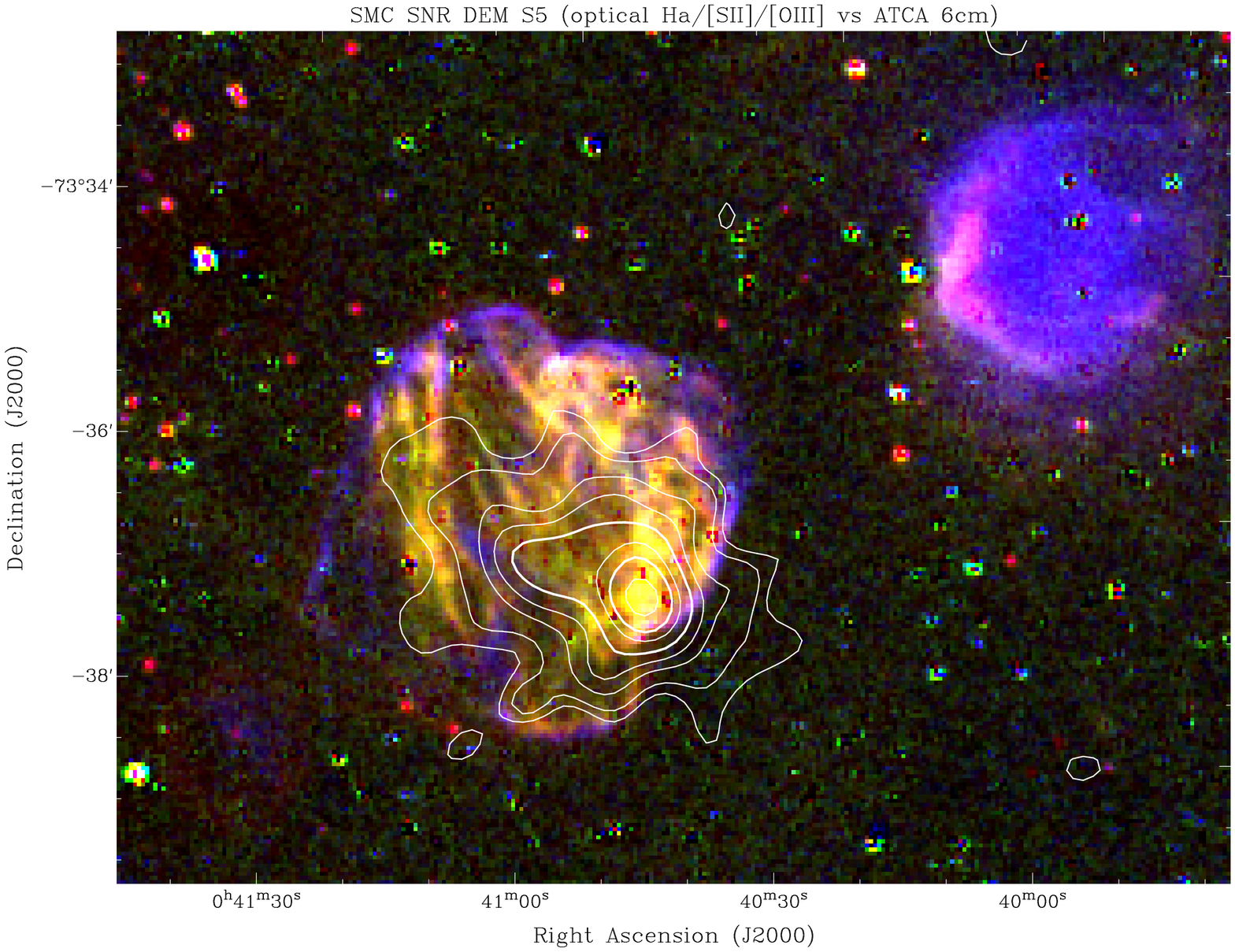}
  \caption{A composite optical image (RGB = \Halpha, \SII\ and \OIII, respectively,
  all continuum-subtracted) of SMC SNR DEM\,S5
  overlaid with (A) \xmm\ broad-band contours and (B) ATCA 6-cm contours.
  The X-ray contours are 15, 20, 25, 30, 35, 40,
  and 50 \expo{-6} ct~s$^{-1}$~pix$^{-1}$, and the
  radio-continuum contours are 2, 3, 5, 7, 10, 15, 20 and 25~mJy/beam.
  Note that the \OIII\ emission is brightest at the outside of the shell, closest to
  the expanding SNR shock, with \Halpha, \SII\ lagging behind where the post-shock
  gas has cooled somewhat.
  The prominent bluish  object at the upper right (north-west --
  centred at RA(J2000)=00$^h$40$^m$02.6$^s$ and DEC(J2000)=--73\D33\arcmin53\arcsec),
  with faint \OIII\ inside and \Halpha\ around much of the ill-defined
  periphery, is  a photo-ionised region. Another, much fainter region of \OIII\ emission is just
  south-east of DEM\,S5 (RA(J2000)=00$^h$41$^m$32.7$^s$ and
  DEC(J2000)=--73\D38\arcmin22\arcsec).}
  \label{dems5}
 \end{center}
\end{figure*}

The remnant is $\sim$210\arcsec\ (60\, pc) in size, with a complex
shell-like optical structure shown in Fig.~\ref{dems5} (RGB=\Halpha/\SII/\OIII, 
all with a matched continuum subtracted). The optical shell could 
comprise of two intersecting ring systems,
one elongated north-south, and the other (most prominent in \OIII)
elongated east-west. However, we acknowledge that this needs to be 
confirmed by more precise kinematic analysis. The outer periphery 
of almost the entire shell is marked
by \OIII\ emission, characteristic of excitation by shocks with $v\gtrsim 100\,{\rm km\; s^{-1}}$,
with stronger \Halpha\ and \SII\ emission from the cooling regions behind the shock.
Our new \xmm\ image (Fig.~\ref{dems5})
shows multiple peaks around and within the optical shell.
However, both the X-ray and the
radio-continuum images have a slightly smaller extent than that seen optically,
most likely due to lower resolution and sensitivity. Nevertheless,
DEM\,S5 ranks among the largest SNRs in the SMC.
EPIC spectra of the remnant were extracted for PN and MOS2 (Fig.~\ref{fig-snr-spectra}a)
and are typical of SNRs; the MOS1 detector did not cover the source.

We note that the peak emission in the X-ray and radio-continuum images
coincides with a small region, $\sim$16\arcsec\ in size and centred at
RA(J2000)=00$^h$40$^m$47.7$^s$ and DEC(J2000)=\hbox{--73\D37\arcmin03\arcsec},
where the \Halpha\ and \SII\ emission is also brightest.
This may simply be a dense cloud that has been overrun by the SNR shock,
but the X-ray peak is not spatially resolved and has a harder spectrum than the
rest of the remnant, so we cannot exclude a line-of-sight association
with an active galaxy or quasar.

\subsection{The SMC SNR\,B0050$-$72.8 (DEM\,S68; HFPK\,285; J005240-723820)}

The complex nebulae SNR\,B0050$-$72.8 was identified as a SNR by
\citet{1984ApJS...55..189M} and \citet{1982MNRAS.200.1007M} based on
optical and radio-continuum (MOST) detections. They suggest this
remnant has a steep (non-thermal) radio-continuum spectrum and is
most likely an older SNR since it has a diameter of $\sim$57~pc.
\citet{2005MNRAS.364..217F} confirmed its radio non-thermal nature
($\alpha$=--1.0) and \citet{2007MNRAS.376.1793P} estimated a
[S\,{\sc ii}]/H$_\alpha$ ratio of 0.4. A {\sl ROSAT} detection was
reported by \citet{2000A&AS..142...41H} who designated it HFPK\,285.

\begin{figure*}
 \begin{center}
 \includegraphics[angle=-90,width=88mm]{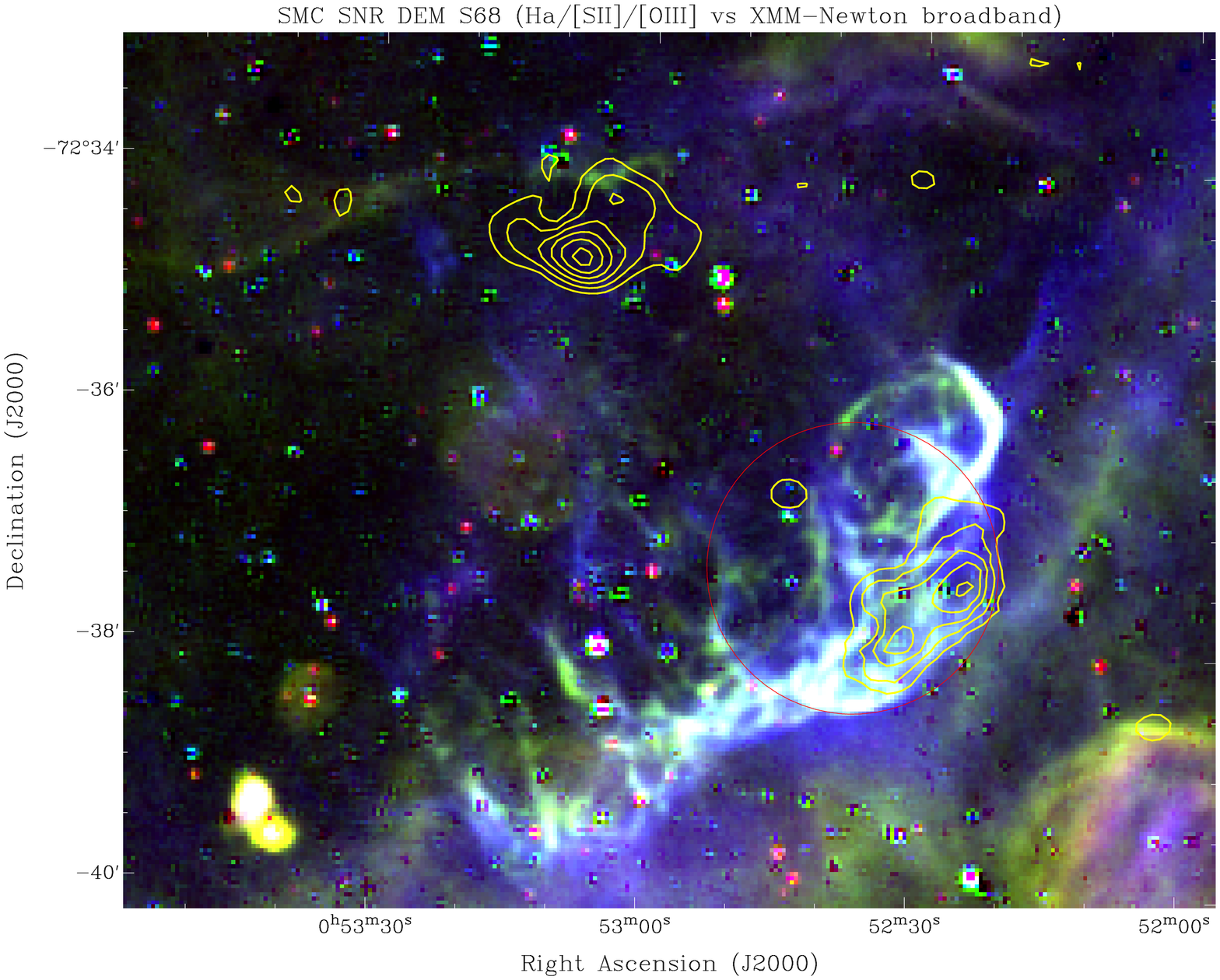}
 \includegraphics[angle=-90,width=88mm]{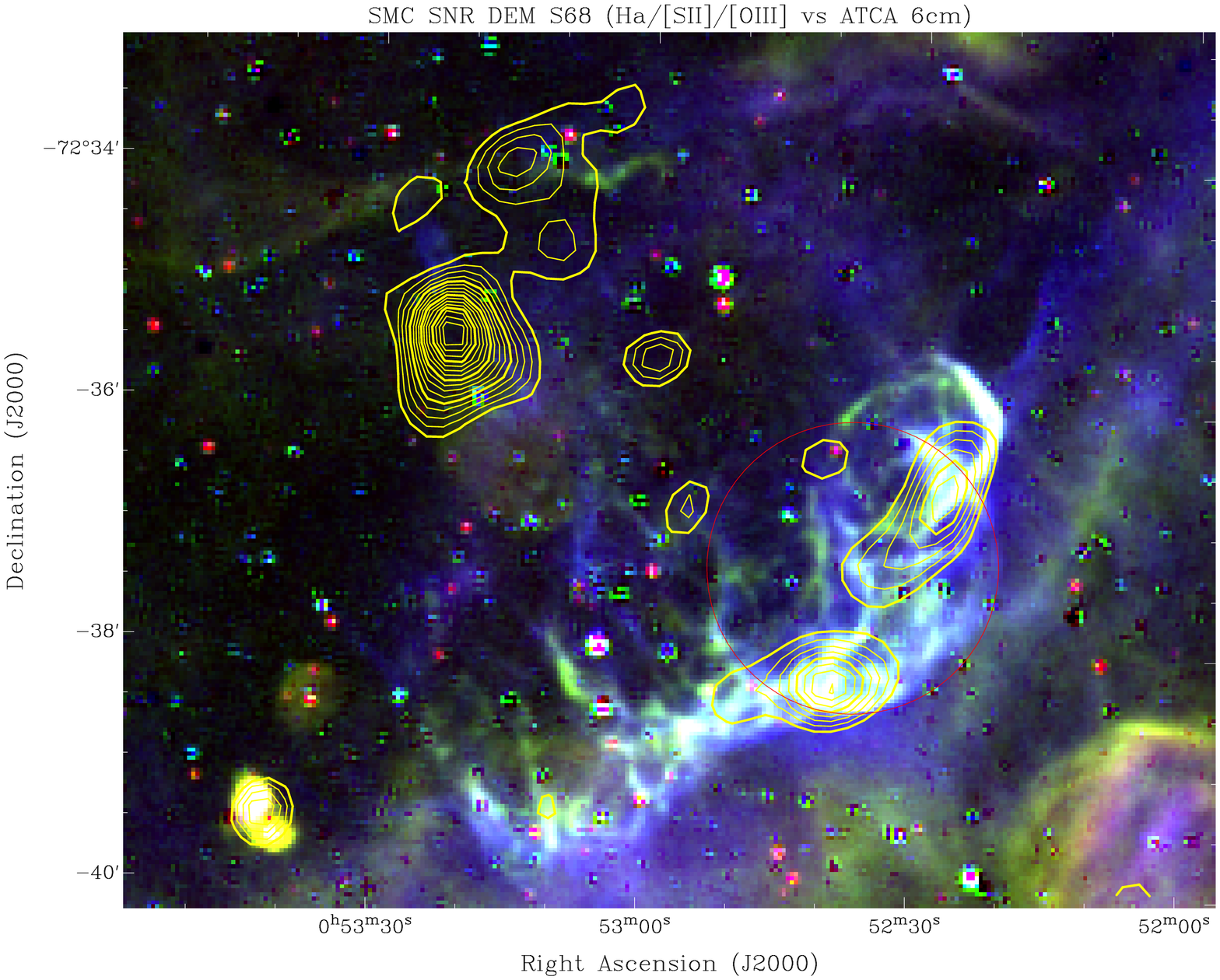}
  \caption{A composite optical image (RGB -- \Halpha, \SII\ and \OIII) of SMC SNR\,B0050$-$72.8
  overlaid with (A) \xmm\ broad-band contours and (B) ATCA 6-cm contour map.
  X-ray contours are 15, 20, 25, 30, 35, and 40 \expo{-6} ct~s$^{-1}$~pix$^{-1}$.
  Radio-continuum contours are from 1 to 5~mJy/beam in steps of 0.25~mJy/beam. The red circle
  indicates the position and extent of the remnant. The bright object in
  the lower left corner is the compact \HII\ region DEM\,S69.}
 \label{dems68}
 \end{center}
\end{figure*}

From our optical images we observe a number of ring structures, with 
a distinctive loop, that may correspond to the radio and X-ray emissions, 
centred at RA(J2000)=00$^h$52$^m$36.9$^s$ and
DEC(J2000)=--72\D37\arcmin18.5\arcsec. The remnant's optical
emission is brighter to the south and west with no emission in the
north-east region. Two bright radio-continuum peaks
(Fig.~\ref{dems68}) indicate radio emission which coincides with the
optical peaks.

We estimate SNR\,B0050$-$72.8's diameter as 145$\pm$10\arcsec\
(42~pc), somewhat smaller than suggested by
\citet{1984ApJS...55..189M}. This is most likely due to the
complexity of the region which is superimposed by a
$\sim$300\arcsec\ (87~pc) elongated shell (or bubble).

SNR\,B0050$-$72.8 was covered by three \xmm\
observations, in all cases at the rim of the EPIC FOV. The EPIC
image shows extended emission at the south-west rim of the optical
emission. However, the peak X-ray emission doesn't coincide with
either the radio-continuum or the optical emission. The EPIC image
also shows extended emission in the north-east, about 1\arcmin\
outside of the optical emission. The similar X-ray colours of the
two emission components suggest they are either both related to the 
larger SNR (bubble) or that the north-east source is an independent 
new SNR (Fig.~\ref{dems68}). We could not find any obvious 
radio-continuum or optical emission corresponding to this X-ray source.

An alternate explanation is an SNR in a larger cavity or bubble. 
In this model, the radio and X-ray to the NE and in the red circle 
may just be the brightest peaks in a larger shell of emission. 
This larger bubble in the centre of Fig.~\ref{dems68}.

\subsection{The SMC SNR\,B0058$-$71.8 (DEM\,S108; HFPK\,45; J010024-713336)}
\begin{figure}
 \begin{center}
 \includegraphics[angle=-90,width=88mm]{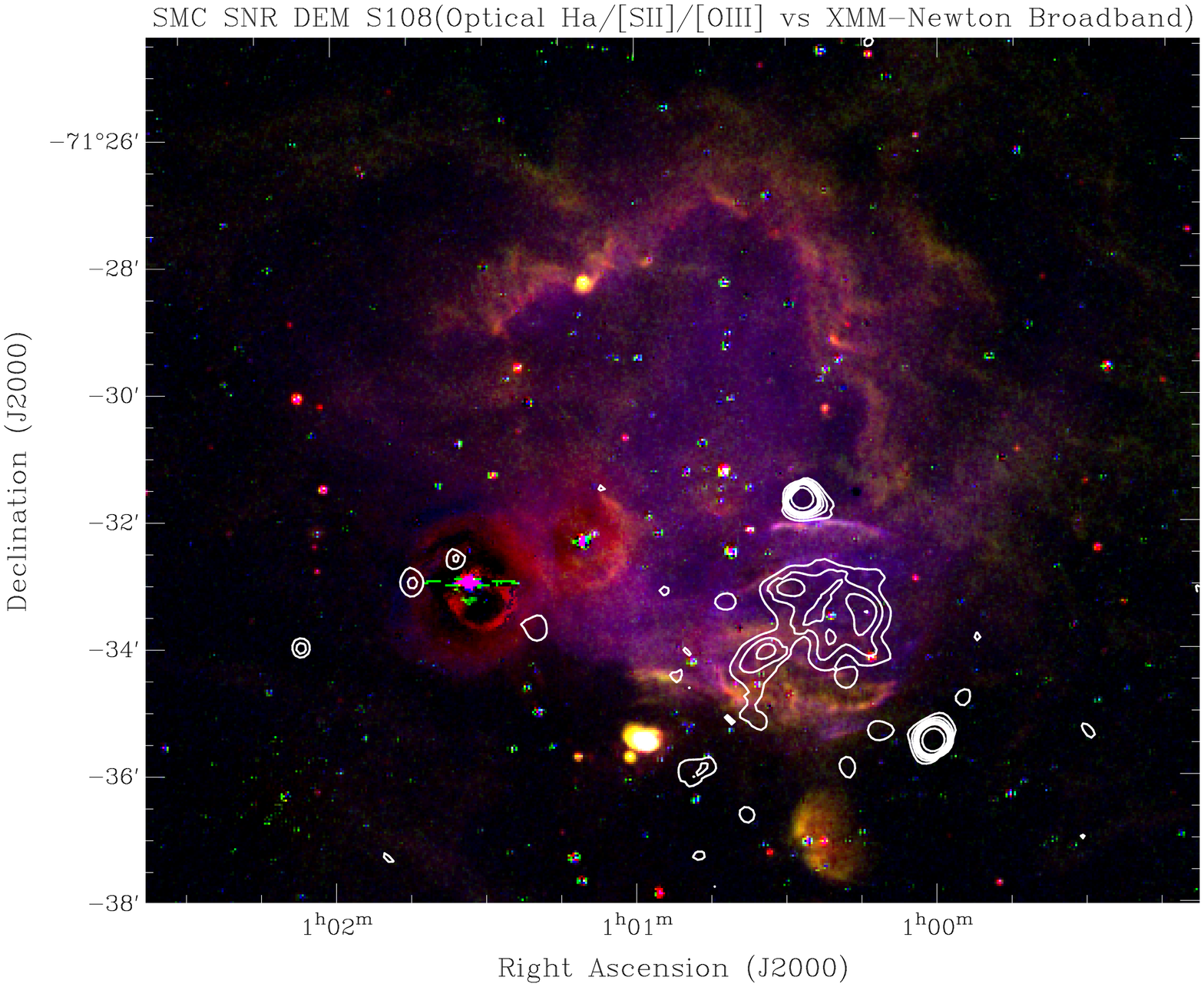}
 \includegraphics[angle=-90,width=88mm]{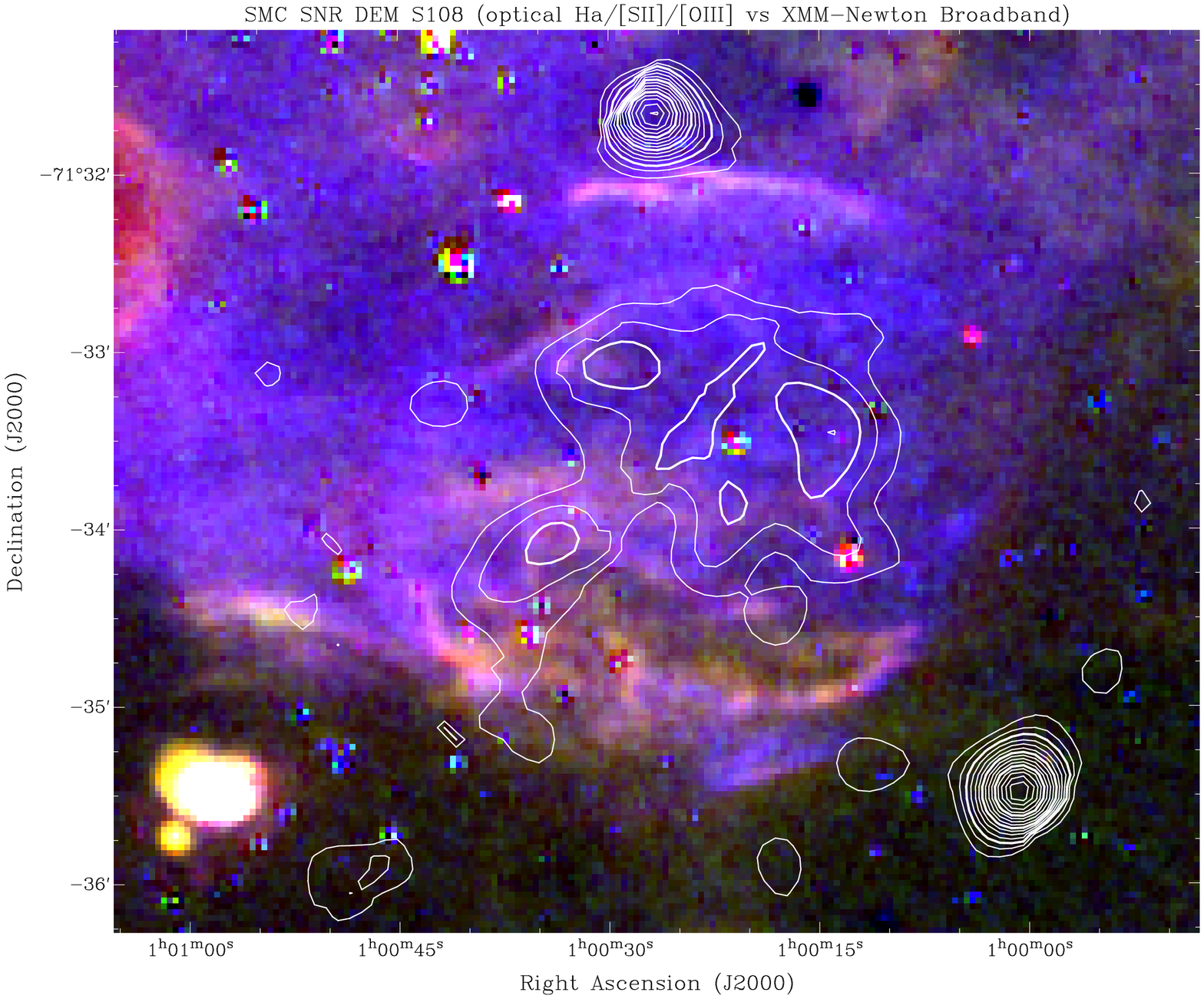}
 \includegraphics[angle=-90,width=88mm]{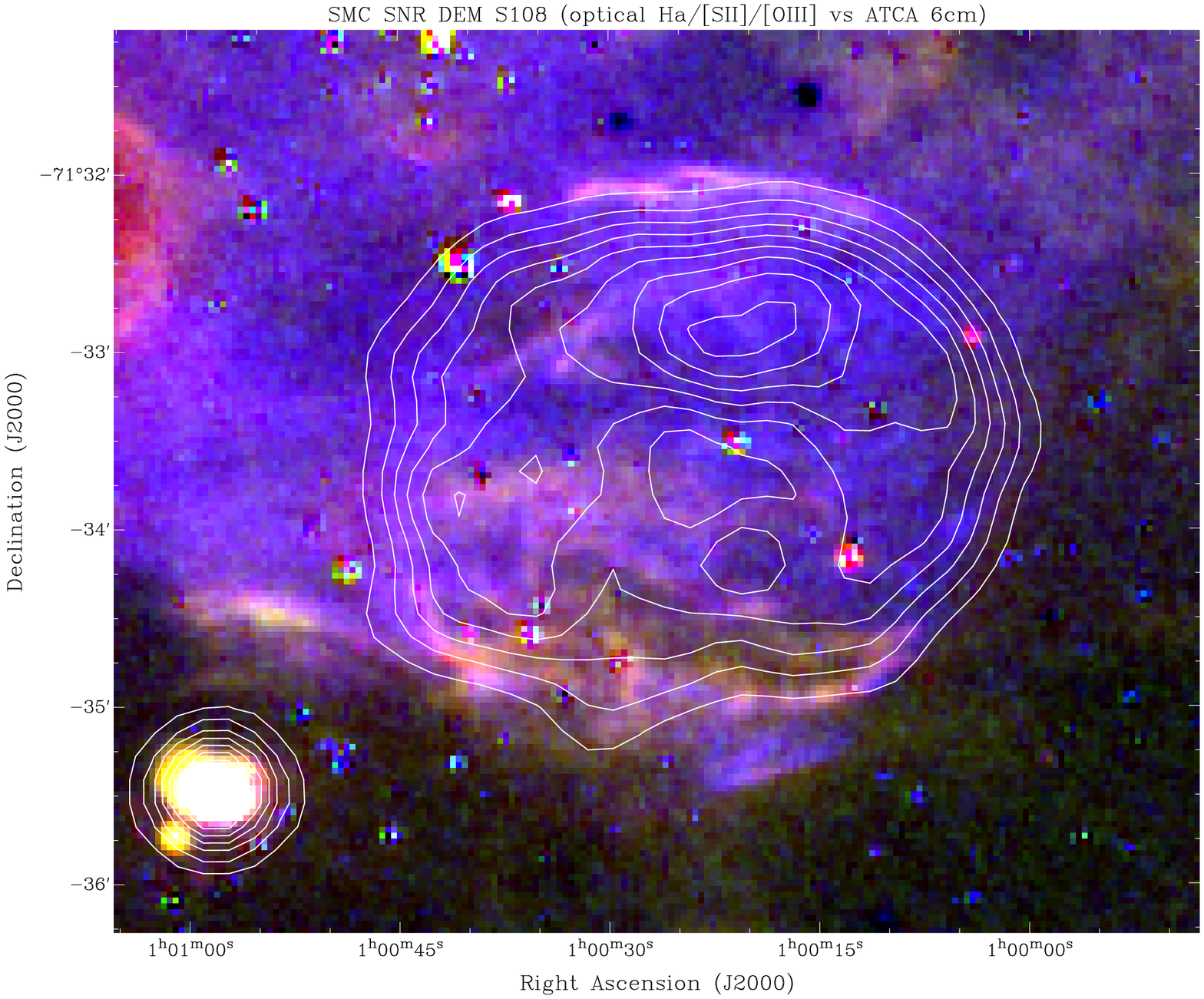}
  \caption{A composite optical image (RGB -- \Halpha, \SII\ and \OIII) of SMC
  SNR\,B0058$-$71.8 overlaid with (A) \xmm\ broad-band contours, showing the entire DEM S108 shell. 
  An inset showing just the remnant overlaid with (B) \xmm\ broad-band contours and (C) ATCA 6-cm
  contours. X-ray contours are 30, 35, 40, 45, 50, 70, and
  150 \expo{-6} ct~s$^{-1}$~pix$^{-1}$. Radio-continuum contours are 0.5, 1, 1.5, 2, 2.5, 3, 3.5, 4, 	and 5~mJy/beam. The bright object in the lower left corner is the compact \HII\ region DEM\,S109.}
 \label{dems108}
 \end{center}
\end{figure}

This remnant was originally classified as an SNR by
\citet{1984ApJS...55..189M} and \citet{1982MNRAS.200.1007M}, based
on both optical and radio-continuum (MOST) detections. It is centered at
RA(J2000)=01$^h$00$^m$24$^s$,
DEC(J2000)=--71\D33\arcmin36\arcsec, near the southwest rim of
the giant \HII\ region DEM\,S108 which may be associated with the stellar
cluster Br\"uck 101 \citep{1976ORROE...1.....B}.
Optical images (Fig. ~\ref{dems108}) show a well-defined, possibly barrel-like shell
with the axis oriented near east-west. Diffuse emission, primarily \OIII, is present throughout
the interior, but this is part of the much larger DEM\,S108 shell, which may or may
not be physically associated with the SNR.
The bright star located very near the (projected) SNR
center is Tyc 9138-1786-1, a foreground star in the Galaxy \citep{2000A&A...355L..27H}.

The 6-cm ATCA radio image shows a very well-defined
elliptical shell for the SNR, measuring $230\arcsec \times 180\arcsec$\ ($67\times 52$\, pc)
\footnote{We estimate that the error in the diameter is $<$10\arcsec\ (3\,pc)},
that corresponds extremely well with the optical filaments. Our earlier radio
studies \citep{2005MNRAS.364..217F} confirmed its non-thermal nature, with 
spectral index $\alpha$=--0.8). \citet{2000A&AS..142...41H}
detected this source with the {\sl ROSAT} PSPC (HFPK\,45).

EPIC observations of SNR\,B0058$-$71.8 included 11.9~ks (PN) and
13.5~ks (MOS) exposures. From the resulting image, we found its
X-ray diameter to be $150\arcsec\times120$\arcsec\ ($43\times35$~pc);
significantly smaller than its optical and radio extent.
We note that a higher background emission, when considered with its 
apparent older age\footnote{X-ray emission from SNRs tend to fade 
earlier since it is predominantly thermal.} based on its X-ray spectrum 
(Fig.~\ref{fig-snr-spectra}c) and large optical/radio diameter, may account 
for this. Although a number of X-Ray SNRs are centrally condensed
(like this object), most have clear shells at radio and optical frequencies.
Paradoxically, its \OIII\ dominated emission confuses this picture,
since these SNRs tend to be younger. However, it is possible that the
\OIII\ that fills the SNR is part of the much larger DEM S108 shell.

We used the \xmm\ observation having the longest exposure
time, to extract an EPIC-PN spectrum (Fig.~\ref{fig-snr-spectra}c).
This covered only the western portion of two emission regions.

\subsection{The SMC SNR candidate HFPK\,334 (J010329-724723)}

\begin{figure}
 \begin{center}
 \includegraphics[angle=-90,width=88mm]{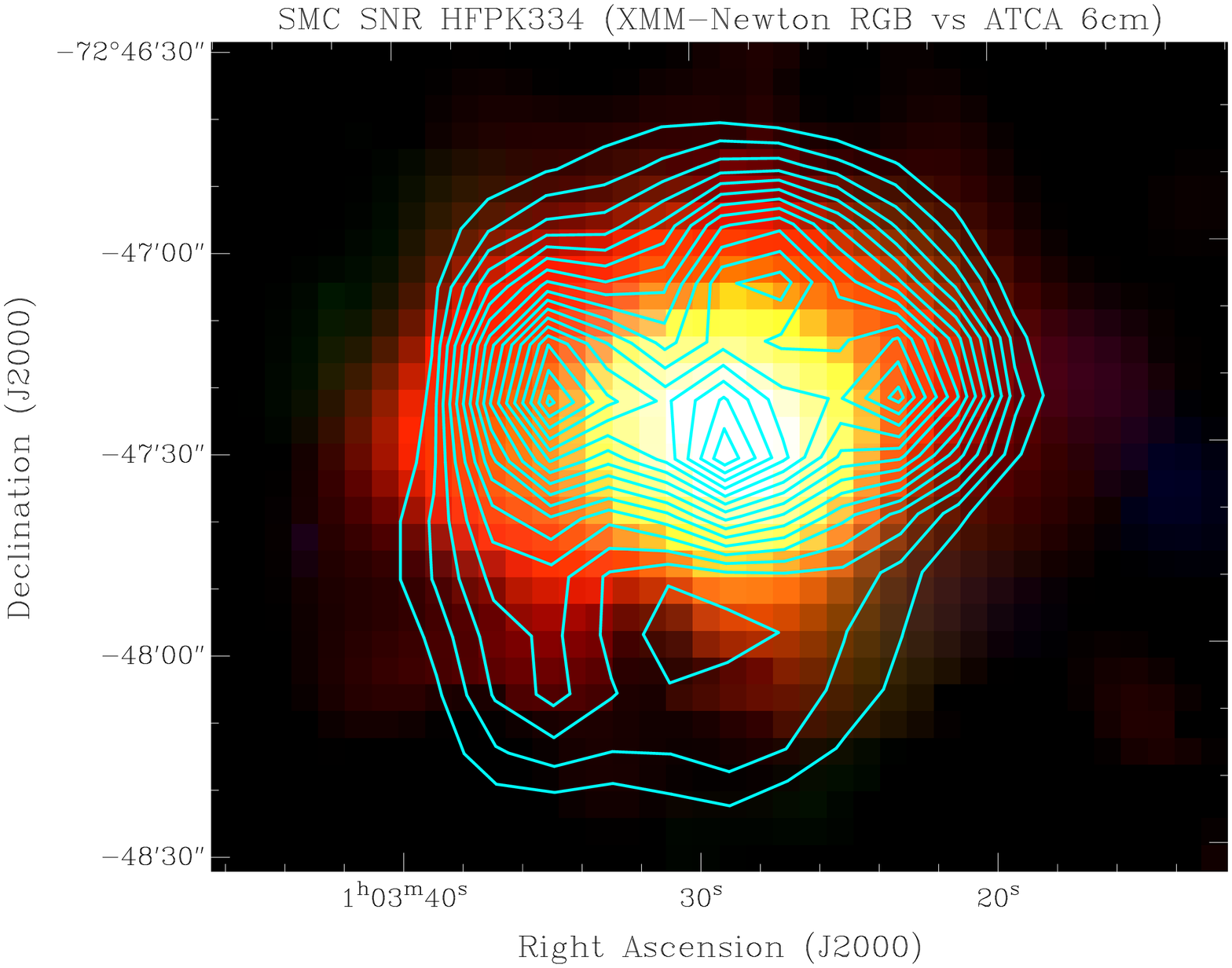}
 \includegraphics[angle=-90,width=68mm]{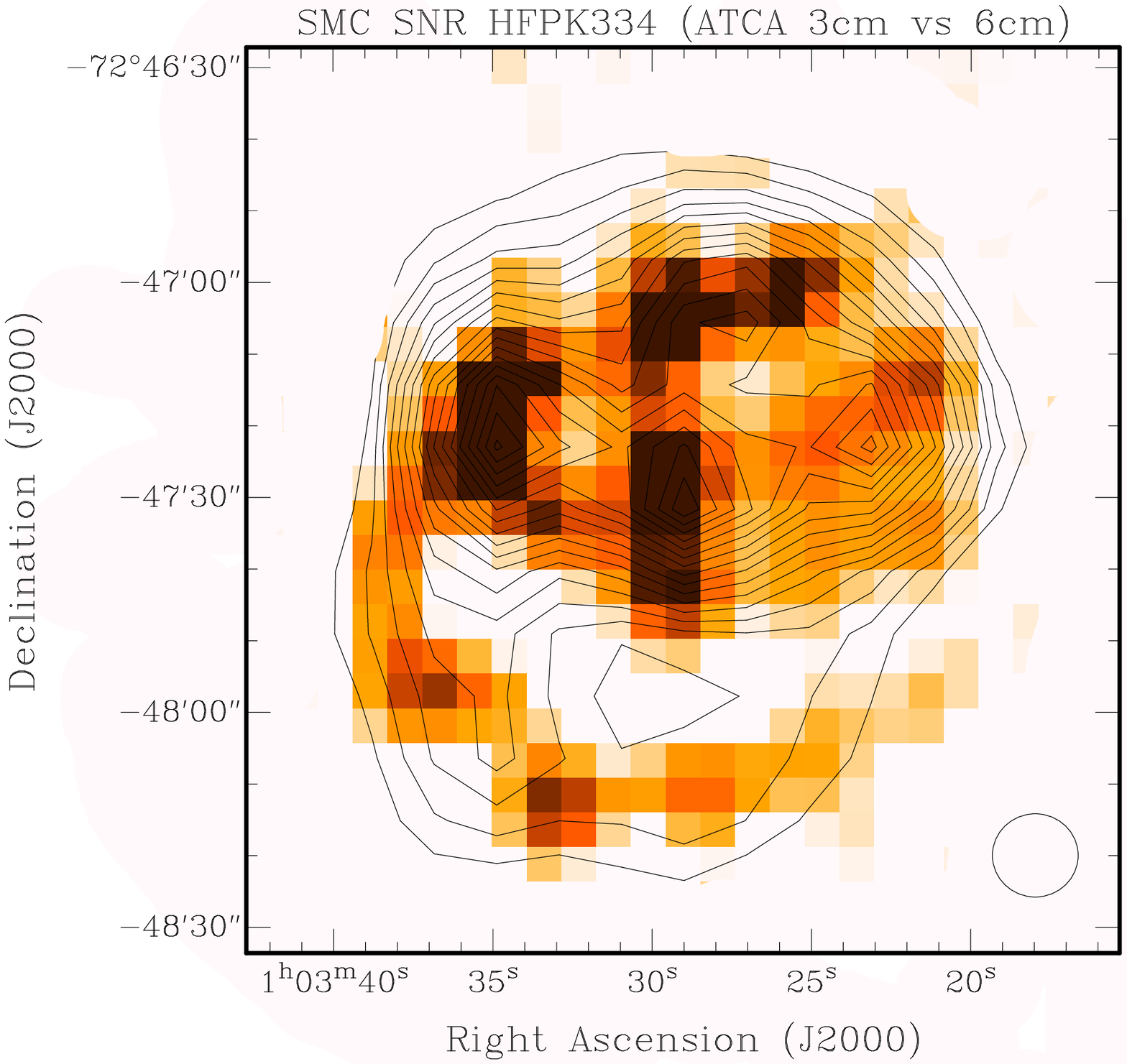}
  \caption{(A) A composite \xmm\ image (RGB -- soft (0.2-1.0~keV),
  medium (1.0-2.0~keV) and hard (2.0-4.5~keV)) of the SMC SNR\,HFPK\,334
  overlaid with ATCA 6-cm contour map. 
  (B) ATCA 3-cm  image overlaid with ATCA 6-cm contour map. The black
  circle represents the 3-cm beam size of 12 $\times$ 11\arcsec. 
  The radio-continuum contours at 6-cm are from 0.27 to 3.0~mJy/beam in 
  0.13~mJy/beam steps.}
 \label{hfpk334}
 \end{center}
\end{figure}

We found a new candidate SNR based on its X-ray colour or hardness
(Fig.~\ref{hfpk334}). Listed as number 334 in the PSPC catalogue
\citep{2000A&AS..142...41H}, it co-identifies with an extended
13/6~cm radio source \citep{2000A&AS..142...41H}. Following the
naming convention used by \citet{2004A&A...421.1031V} for SNRs
detected in {\sl ROSAT} data, we designate this candidate SNR
HFPK\,334. Its \mbox{X-ray} spectrum, shown in Fig.~\ref{fig-snr-spectra}d,
was extracted using data from all three EPIC instruments.

Measured overall (total) integrated radio-continuum flux densities
are 27.7, 22.1, 18.0, 14.3, and 8.6~mJy\footnote{An error is
estimated to be $<$10\%.} at 36, 20, 13, 6, and
3~cm, respectively. Using these values, we find a typical SNR
spectral index ($\alpha$) of \hbox{$-0.5\pm0.1$}. However, upon closer
examination the total spectral index appears to be composed of two
components, one from 36 to 6~cm with a value ($\alpha_1$) of
$-0.4\pm0.1$. The second is from 6 to 3~cm with a value
($\alpha_2$) of $-0.9\pm0.1$. At 6 and 3~cm, the source is
resolved and extended with a diameter of $100\arcsec \times 80\arcsec$
($29 \times 23.2$~pc) with an estimated error of $\pm10\arcsec$ ($\pm$2.9~pc). 
From Fig.~\ref{hfpk334} (at 3~cm) we can see a
bright point source in the centre of the proposed remnant. We
measured the flux of this point source alone to be identical at 6
and 3~cm (2.4$\pm$0.4~mJy). This suggests that HFPK\,334 central
source is associated with the SNR shell, indicating a possibility of
this object being a Pulsar Wind Nebula (PWN). However, adding the longer 
baselines from the 6~km array (January 2008 observations) resolved out
the SNR's extended emission and no point source could be associated with 
the remnant. The XMM-Newton data also shows an extended and possibly non-thermal 
source with a bright central peak, a similar situation to LMC SNR~B0453-685
\citep{2003ApJ...594L.111G}. Also, the two component flat radio
spectral index of this point source ($\alpha$=--0.0) seems to
support this. Finally, we note that no reliable polarisation was 
detected at either 6 or 3~cm to a level of 10\%.

This object is thus unusual, as it is the first suspected PWN in the
SMC. Furthermore, even very deep optical observations with various
filters ([O$_\mathrm{III}$], [S$_\mathrm{II}$] and H$_\alpha$)
failed to detect any optical emission from this proposed
SNR. This would be the rare case of a SNR in the SMC without
optical identification which even further indicate the complex nature of
this object. Further high-resolution radio-continuum and X-ray studies would
help to understand this object.

\subsection{SMC SNRs not detected in X-ray surveys}

Two objects classified as `definite' SMC SNRs are not detected in
any X-ray surveys: SNR\,J004728-730601 \citep{2001AJ....122..849D}
and SNR\,J005800-721101 \citep{2006MNRAS.367.1379R}. In spite of the
fact that both were covered by  several  \xmm\ pointings, we
could not detect any distinctive\footnote{SNR\,J004728-730601 may
exhibit a very low level X-ray emission; future \xmm\ observations
may reveal the true nature of this source.} X-ray emission.

There are several other sources in the SMC classified as SNR
candidates. These include J0051.9-7310 suggested by
\citet{2001AJ....122..849D} and J011333-731704 suggested by
\citet{2005MNRAS.364..217F}, both based on radio-continuum images
and tentative {\sl ROSAT} detections. Although these regions contain
point-like X-ray emission that is suspicious of background objects, 
we did not find any extended X-Ray emission to confirm these objects 
are SNRs. A very thin \SII\ ring surrounding J011333-731704, possibly 
indicative of shocked regions, was the only optical emission detected.

Lack of X-ray emission from these SNR candidates is relatively surprising 
since some X-ray emission is generally expected. We acknowledge that this 
is a very complex issue, since we are dealing with the limiting 
sensitivities of very different instruments and at very different wavelengths 
coupled with observational parameters (e.g. foreground extinction and its 
variable effect on different wavelengths)  and the variable nature of the 
sources being observed (large, evolved SNRs versus younger, smaller objects or 
objects evolving in diffuse fairly uniform ISM versus impacting nearby clumpy material).
\citet{2004A&A...425..443P} and \citet{2007AJ....133.1361P} found a
number of optical/radio SNRs in NGC\,300, M\,81, M\,101, NGC\,2403,
NGC\,4736, NGC\,6946 without any X-ray trace. They argue that the
local ambient density plays a major role in addition to instrument
sensitivity in these more distant galaxies.

Stupar et al. (in preparation) found 21 new Galactic SNRs using high
resolution optical detection and spectroscopy, from the AAO/UKST SuperCOSMOS 
\Halpha\ Survey \citep{2005MNRAS.362..689P}. About half (10) of these could 
not be correlated with any \hbox{X-ray} or radio-continuum emission. They 
conclude that the local environment into which an individual SNR is propagating 
plays a very important role in this. Other factors include selection effects 
based on supernova type and survey sensitivity.

\subsection{SMC SNRs diameter distribution}

With the new and moderate resolution surveys at various radio frequencies
we can measure more precisely the diameters of the SMC SNRs and SNR
candidates. We estimate the precision of our diameter measurements to be 
$<$2\arcsec\ ($<$0.6~pc). Despite that some SMC SNRs (for example SNR~B0058-71.8)
exhibit different diameters across different part of the spectrum,
we find that the majority of SNRs have consistent diameters at various
frequencies.

\begin{figure}
 \begin{center}
 \includegraphics[width=88mm]{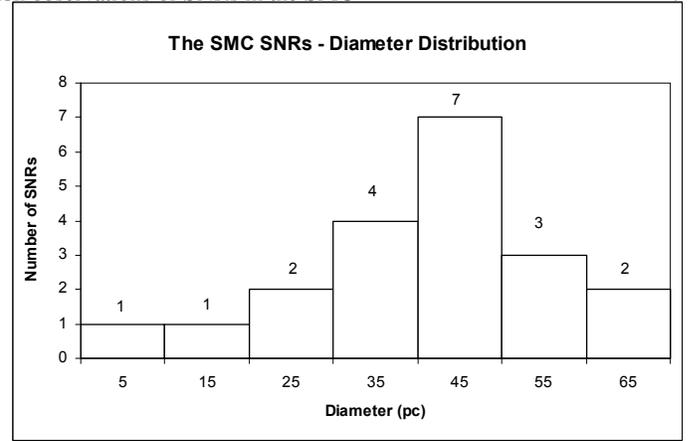}
  \caption{A diameter distribution of the SMC SNRs and SNR candidates.
  We assume a distance to the SMC of 60~kpc.}
 \label{smcd}
 \end{center}
\end{figure}

In Table~\ref{smcsnrs} (Col.~5) are a list of diameters for all SMC SNRs
and SNR candidates. Fig.~\ref{smcd} shows that the majority (7 out
of 20) of the SMC SNRs have an average diameter of $\sim$45~pc. That
indicates that most of the SMC SNRs might be in the adiabatic phase
of evolution.

\section{Conclusions}
 \label{conclusions}

We have analysed our new and archival \xmm\ observations in the direction to
the SMC, discovering one new X-ray SNR candidate (HFPK\,334) with strong
indication for the existence of a PWN. In addition, we have examined three
previously known SNRs (DEM\,S5, SNR\,B0050$-$72.8, and SNR\,B0058$-$71.8), 
expanding their X-ray emission details significantly. More specifically we found:
\begin{enumerate}
\item HFPK\,334 as a new bona-fide SNR, without an optical signature.
\item DEM\,S5 as a definite SNR with a very complex shell structure.
\item SNR\,B0050$-$72.8 as either an SNR or a supernova which exploded in a 
large bubble with a complex filamentary structure.
\item SNR\,B0058$-$71.8 an SNR with significantly smaller X-ray diameter not 
seen in any other known SNR.
\end{enumerate}

All known SNRs in the SMC show a vast variety of types and environments in 
which they exist. Some SNRs could be confused with bubbles due to very complex 
and crowded fields. We find that the majority of the SMC's SNRs are in the 
adiabatic phase of their evolution with an average diameter of 45~pc.

We argue that high-resolution systematic X-ray, optical and
radio-continuum observations of the SMC/LMC and other galaxies allow
us to understand the overall multi-wavelength properties of SNRs.

\begin{acknowledgements}
The \xmm\ project is supported by the Bundesministerium f\"ur
Wirtschaft und Technologie/Deutsches Zentrum f\"ur Luft- und
Raumfahrt (BMWI/DLR, FKZ 50 OX 0001) and the Max-Planck Society.
We used the Karma/MIRIAD software packages
developed by the ATNF. The Australia Telescope Compact Array is part
of the Australia Telescope which is funded by the Commonwealth of
Australia for operation as a National Facility managed by the CSIRO.
MDF would like to thank the Max-Planck-Institut f\"ur
extraterrestrische Physik for support. PFW acknowledges the many contributions
of R.C. Smith of CTIO and other members of the MCELS team, and financial support
from U.S. National Science Foundation through grant AST-0307613.
BER acknowledges support from the U.S. National Science Foundation through grant
AST-0353997 to the Keck Northeast Astronomy Consortium REU program. We thank the
referee for their excellent comments that improved this manuscript.
\end{acknowledgements}

\bibliographystyle{aa}
\bibliography{9642mcs}

\end{document}